\documentclass[reprint,twocolumn,showpacs,preprintnumbers,amsmath,amssymb,superscriptaddress,aps,prl]{revtex4-1}
\usepackage[maccyr]{inputenc}
\usepackage[T2A]{fontenc}
\usepackage{bm}
\usepackage{epsfig,amssymb,amsmath,bm}
\usepackage[usenames,dvipsnames,svgnames,table]{xcolor}

\begin{document}

\title{The universal expression for the amplitude square in quantum electrodynamics}
\author{K. S. Karplyuk}
\email{karpks@hotmail.com}
\affiliation{Department of Radiophysics, Taras Shevchenko University, Academic Glushkov prospect 2, building 5, Kyiv 03122, Ukraine}
\author{O. O. Zhmudskyy}
\email{ozhmudsky@physics.ucf.edu}
\affiliation{Department of Physics, University of Central Florida, 4000 Central Florida Blvd. Orlando, FL, 32816 Phone: (407)-823-4192}
\begin{abstract}

The universal expression for the amplitude square $|\bar{u}_fMu_i|^2$ for any matrix of interaction $M$ is derived.
It has obvious covariant form. It allows the avoidance of calculation of products of the Dirac's matrices traces and allows
easy calculation of cross-sections of any different processes with polarized
and unpolarized particles.
\end{abstract}

\pacs{12.20.-m}

\maketitle

\section{Introduction}
Amplitude  square $|\bar{u}_fMu_i|^2$ calculations are necessary in order to find probability transactions for any processes
in quantum electrodynamics.  The interac\-tion matrix $M$
is the combination of the Dirac matrices and their products.  This circumstance causes very labor-intensive  calculation even if the Feynman
technique of trace of matrix products calculation is used\cite{f}.  Especially labor-intensive calculations are when polarization of in- and out- particles is taken into account.
That is why the such calculations often do not take particle polarization into account.
Usually for each particular process $|\bar{u}_fMu_i|^2$ is calculated separately. There are very many papers devoted to calculation of
$|\bar{u}_fMu_i|^2$ for a particular processes.

However, all interaction matrices have the same structure and set of permissible
matrices is restricted. Any $4\times 4$  matrix  can be represented as
\begin{equation}
M=I\hat{1}+V_\alpha\gamma^\alpha+W_\alpha\pi^\alpha+\frac{1}{2}F_{\alpha\beta}\sigma^{\alpha\beta}+J\hat{\iota}.
\end{equation}
Here $\hat{1}$ --- unit matrix, $\gamma^\alpha$ --- four Dirac's matrices, $\hat{\iota}=\gamma^0\gamma^1\gamma^2\gamma^3$, $\pi^\alpha=\gamma^\alpha\hat{\iota}$, $\sigma^{\alpha\beta}=\frac{1}{2}(\gamma^\alpha\gamma^\beta-\gamma^\beta\gamma^\alpha)$, $I$ and $J$ scalar and pseudoscalar, $V_\alpha$ and $W_\alpha$ vector
and pseudo-vector, $F_{\alpha\beta}$ anti-symmetrical tensor.

In- and out- fermions are represented by Dirac's bispinors of the same type:
\begin{equation}
u=\sqrt{\frac{p_o+mc}{2}}\Bigl(\hat{1}+\frac{\bm{p}\bm{\varsigma}_1}{p_0+mc}\Bigr)\frac{\hat{1}+i\bm{n}\bm{\varsigma}_2}{\sqrt{2(1+n_z)}}
\left[\!\!\begin{array}{c}
1\\0\\
0\\0
\end{array}\!\!\right].
\end{equation}
Here $\bm{p}\bm{\varsigma}_1=p_x\sigma^{01}+p_y\sigma^{02}+p_z\sigma^{03}$, $\bm{n}\bm{\varsigma}_2=n_x\sigma^{23}+n_y\sigma^{31}+n_z\sigma^{12}$,
$\bm{n}$ --- three dimensional unit spin pseudo-vector in particle's own reference frame, $n_0=0,n_1=-n_x,n_2=-n_y,n_3=-n_z$.
In particles's own reference frame it's linear momentum is zero.
For the bispinor $u$ the relativistically covariant  normalization $\bar{u}u=mc$ is used.

Thus the possible choices for $|\bar{u}_fMu_i|^2$ are restricted. So, for all of them $|\bar{u}_fMu_i|^2$ can be calculated and a universal expression can be derived.  This expression can be  used for all possible interaction matrices.  Such an expression was derived in \cite{k} but $|\bar{u}_fMu_i|^2$  is expressed through the three dimensional quantities in laboratory reference frame. In most cases it is preferable to have Lorentz's covariant expression which is derived below.

\section{Covariant expression for amplitude square}
Let us write $|\bar{u}_fMu_i|^2$ as $(\bar{u}_fMu_i)(\bar{u}_fMu_i)^*$ and use the equality:
\begin{equation}
(\bar{u}_fMu_i)^*=(\bar{u}_fMu_i)^\dag=u_i^\dag M^\dag \gamma^{0\dag}u_f=u_i^\dag\gamma^0\tilde{M}u_f.
\end{equation}
Here
\begin{equation}
\gamma^0\tilde{M}=M^\dag\gamma^{0\dag}=\gamma^0(I^*\hat{1}+V_\alpha^*\gamma^\alpha-W_\alpha^*\pi^\alpha-\frac{1}{2}F_{\alpha\beta}^*+J^*\hat{\iota}).
\end{equation}
Which leads to:
\begin{equation}
|\bar{u}_fMu_i|^2=(\bar{u}_fMu_i)(\bar{u}_i\tilde{M}u_f)=\mathrm{Sp}\,u_f\bar{u}_fMu_i\bar{u}_i\tilde{M}.
\end{equation}
Let us take into account that for the bispinor (2)
\begin{eqnarray}
u\bar{u}=\frac{mc}{4}\Bigl(\hat{1}+\frac{p_\alpha\gamma^\alpha}{mc} \Bigr)\Bigl(\hat{1}-is_\alpha\pi^\alpha\Bigr)= \nonumber\\
=\frac{mc}{4}\Bigl(\hat{1}+\frac{p_\alpha\gamma^\alpha}{mc} -is_\alpha\pi^\alpha-\frac{i}{2}\varsigma_{\alpha\beta}\sigma^{\alpha\beta}\Bigr).
\end{eqnarray}

Here $s_\alpha$ --- spin pseudo-vector $n_\alpha$ coordinates  in the reference frame where a fermion has momentum $p_\alpha$.
Vector $s^\alpha$ has coordinates
\begin{equation}
s^0=0,\hspace{4mm}s^1=n_x,\hspace{4mm}s^2=n_y,\hspace{4mm}s^3=n_z,\hspace{4mm}\bm{n}\cdot\bm{n}=1
\end{equation}
in the fermion's reference frame, where it is at rest.

Vector $s^\alpha$ has the following coordinates in the reference frame in which the fermion has linear momentum $p_\alpha$
\begin{equation}
s^0=\frac{\bm{n}\cdot\bm{p}}{mc},\hspace{7mm}\bm{s}=\bm{n}+\frac{\bm{p}}{p_0+mc}\frac{\bm{n}\cdot\bm{p}}{mc}.
\end{equation}

Spin tensor
\begin{equation}
\varsigma^{\alpha\beta}=\varepsilon^{\alpha\beta\mu\nu}\frac{p_\mu s_\nu}{mc}
\end{equation}
has coordinates in the fermion's reference frame:
\begin{equation}
\varsigma^{01}=\varsigma^{01}=\varsigma^{01}=0,\hspace{4mm}\varsigma^{23}=n_1,\hspace{4mm}\varsigma^{31}=n_2,\hspace{4mm}\varsigma^{12}=n_3.
\end{equation}
Here $\varepsilon^{\alpha\beta\mu\nu}$ --- entirely anisymmertical tensor,  $\varepsilon^{0123}=1$, the same tensor  $\varepsilon_{0123}=-1$.
Note that
\begin{equation}
\frac{1}{2}\varsigma_{\alpha\beta}\sigma^{\alpha\beta}=\frac{p_\alpha\gamma^\alpha}{mc}s_\beta\pi^\beta.
\end{equation}

For the $|\bar{u}_fMu_i|^2$ with a help of (5) and (6) we have:
\begin{widetext}
\begin{gather}
|\bar{u}_fMu_i|^2=\Bigl(\frac{mc}{4}\Bigr)^2\mathrm{Sp}\Bigl(\hat{1}+\frac{p_{f\alpha}\gamma^\alpha}{mc}\Bigr)\Bigl(\hat{1}-is_{f\alpha}\pi^\alpha\Bigr)
\Bigl(I\hat{1}+V_\alpha\gamma^\alpha+W_\alpha\pi^\alpha+\frac{1}{2}F_{\alpha\beta}\sigma^{\alpha\beta}+J\hat{\iota}\Bigr)\times\nonumber\\
\times\Bigl(\hat{1}+\frac{p_{i\alpha}\gamma^\alpha}{mc}\Bigr)\Bigl(\hat{1}-is_{i\alpha}\pi^\alpha\Bigr)
\Bigl(I^*\hat{1}+V^*_\alpha\gamma^\alpha-W^*_\alpha\pi^\alpha-\frac{1}{2}F^*_{\alpha\beta}\sigma^{\alpha\beta}+J^*\hat{\iota}\Bigr).
\end{gather}
\end{widetext}
This product contains 400 terms. The trace of most of them is zero. Calculations with the rest of the 164 terms leads to:
\begin{widetext}
\begin{gather}
|\bar{u}_fMu_i|^2\Bigl(\frac{2}{mc}\Bigr)^2=\nonumber\\
=\Bigl[\frac{(p_i\!\cdot\!p_f)}{(mc)^2}-(s_i\!\cdot\!s_f)+
\frac{1}{2}\varsigma_{i\alpha\beta}\varsigma_f^{\alpha\beta}+1\Bigr]{\color[named]{Sepia}I}{\color[named]{Sepia}I^*}+\\
+\Bigl[\frac{(p_i\!\cdot\!p_f)}{(mc)^2}-(s_i\!\cdot\!s_f)-
\frac{1}{2}\varsigma_{i\alpha\beta}\varsigma_f^{\alpha\beta}-1\Bigr]{\color[named]{Blue}J}{\color[named]{Blue}J^*}+\\
+\Bigl\{\bigl(s_i^\mu s_f^\nu+\frac{p_i^\mu p_f^\nu}{(mc)^2}-\varsigma_{i\lambda\,.}^{.\mu}\varsigma_f^{\lambda\nu}\bigr)
(\eta_{\mu\alpha}\eta_{\nu\beta}+\eta_{\mu\beta}\eta_{\nu\alpha})+\bigl[1-(s_i\!\cdot\!s_f)-\frac{(p_i\!\cdot\!p_f)}{(mc)^2}+
\frac{1}{2}\varsigma_{i\mu\nu}\varsigma_f^{\mu\nu}\bigr]\eta_{\alpha\beta}+\nonumber\\
+i\frac{(p_f^\mu-p_i^\mu)}{mc}(s_i^\nu+s_f^\nu)\varepsilon_{\mu\nu\alpha\beta}
\Bigr\}{\color[named]{RubineRed}V^\alpha}{\color[named]{RubineRed}V^{\beta*}}+\\
+\Bigl\{\bigl(s_i^\mu s_f^\nu+\frac{p_i^\mu p_f^\nu}{(mc)^2}+\varsigma_{i\lambda\,.}^{.\mu}\varsigma_f^{\lambda\nu}\bigr)
(\eta_{\mu\alpha}\eta_{\nu\beta}+\eta_{\mu\beta}\eta_{\nu\alpha})-\bigl[1+(s_i\!\cdot\!s_f)+\frac{(p_i\!\cdot\!p_f)}{(mc)^2}+
\frac{1}{2}\varsigma_{i\mu\nu}\varsigma_f^{\mu\nu}\bigr]\eta_{\alpha\beta}+\nonumber\\
+i\frac{(p_f^\mu+p_i^\mu)}{mc}(s_i^\nu-s_f^\nu)\varepsilon_{\mu\nu\alpha\beta}\Bigr\}{\color[named]{Orange}W^\alpha}{\color[named]{Orange}W^{\beta*}}+\\
+\Bigl\{\Bigl(\frac{\varsigma_f^{\alpha\beta}\varsigma_i^{\mu\nu}-\tilde{\varsigma}_f^{\alpha\beta}\tilde{\varsigma}_i^{\mu\nu}}{4}-
\varsigma_f^{\beta\mu}\varsigma^{\alpha\nu}_i\Bigr)+\Bigl[(s_f^\beta s_i^\nu +s_f^\nu s_i^\beta )-
\frac{p_f^\beta p_i^\nu +p_f^\nu p_i^\beta }{(mc)^2}-\varsigma_{f\lambda.}^{.\beta}\varsigma^{\lambda\nu}_i\Bigr] \eta^{\alpha\mu}
+\Bigl[1+\frac{(p_i\!\cdot\! p_f)}{(mc)^2}-(s_i\!\cdot\!s_f)\Bigr]\frac{\eta^{\alpha\mu}\eta^{\beta\nu}}{2}+\nonumber\\
+i\Bigl[(\varsigma_f^{\alpha\mu}-\varsigma_i^{\alpha\mu})\eta^{\beta\nu}+
\frac{(p_f^\beta s_{i\lambda}-p_{i\lambda}s_f^\beta)}{mc}\frac{\varepsilon^{\alpha\lambda\mu\nu}}{2}
-\frac{(p_{f\lambda}s_i^\nu-p_i^\nu s_{f\lambda})}{mc}\frac{\varepsilon^{\alpha\beta\mu\lambda}}{2}\Bigr]\Bigr\}
{\color[named]{ForestGreen}F_{\alpha\beta}}{\color[named]{ForestGreen}F^*_{\mu\nu}}+\\
+\frac{1}{mc}\varsigma_{f\alpha\beta}s^\alpha_i p^\beta_i({\color[named]{Sepia}I}{\color[named]{Blue}J^*}+{\color[named]{Blue}J}{\color[named]{Sepia}I^*})
+i\frac{1}{mc}[(p_f\!\cdot\!s_i)-(p_i\!\cdot\!s_f)]({\color[named]{Sepia}I}{\color[named]{Blue}J^*}-{\color[named]{Sepia}I^*}{\color[named]{Blue}J})+\\
+\frac{(p_f\!\cdot\!s_i)s_{f\alpha}+(p_i\!\cdot\!s_f)s_{i\alpha}+\bigl[1-(s_i\!\cdot\!s_f)\bigr](p_{i\alpha}+p_{f\alpha})}{mc}
({\color[named]{RubineRed}V^\alpha}{\color[named]{Sepia}I^*}+{\color[named]{RubineRed}V^{\alpha*}}{\color[named]{Sepia}I})
+i\frac{\varsigma_{f\alpha\beta}p_{i}^{\beta}-\varsigma_{i\alpha\beta}p_{f}^{\beta}}{mc}
({\color[named]{RubineRed}V^\alpha}{\color[named]{Sepia}I^*}-{\color[named]{RubineRed}V^{\alpha*}}{\color[named]{Sepia}I})+\\
(\varsigma_{f\alpha\beta}s_i^\beta-\varsigma_{i\alpha\beta}s_f^\beta)({\color[named]{Orange}W^{\alpha}}{\color[named]{Sepia}I^*}+
{\color[named]{Orange}W^{\alpha*}}{\color[named]{Sepia}I})+
i\Bigl\{\frac{(p_f\!\cdot\!s_i)p_{i\alpha}+(p_i\!\cdot\!s_f)p_{f\alpha}}{(mc)^2}-\Bigl[1+\frac{(p_i\!\cdot\!p_f)}{(mc)^2}\Bigr]
(s_{i\alpha}+s_{f\alpha})\Bigr\}({\color[named]{Orange}W^\alpha}{\color[named]{Sepia}I^*}-
{\color[named]{Orange}W^{\alpha*}}{\color[named]{Sepia}I})+\\
+\Bigl[\varsigma_f^{\alpha\lambda}\varsigma^{\beta\,.}_{i\,.\lambda}-s_f^\alpha s_i^\beta+\frac{p_f^\alpha p_i^\beta}{(mc)^2}\Bigr]
({\color[named]{ForestGreen}F_{\alpha\beta}}{\color[named]{Sepia}I^*}+{\color[named]{ForestGreen}F^*_{\alpha\beta}}{\color[named]{Sepia}I})
+i(s_{f\nu}+s_{i\nu})\frac{p_{f\mu}+p_{i\mu}}{mc}\frac{\varepsilon^{\alpha\beta\mu\nu}}{2}
({\color[named]{ForestGreen}F_{\alpha\beta}}{\color[named]{Sepia}I^*}-{\color[named]{ForestGreen}F^*_{\alpha\beta}}{\color[named]{Sepia}I})+\\
(\varsigma_{i\beta\alpha}s_f^\beta+\varsigma_{f\beta\alpha}s_i^\beta)({\color[named]{RubineRed}V^\alpha}{\color[named]{Blue}J^*}
+{\color[named]{RubineRed}V^{\alpha*}}{\color[named]{Blue}J})
+i\Bigl\{\frac{(p_f\!\cdot\! s_i)p_{i\alpha}-(p_i\!\cdot\! s_f)p_{f\alpha}}{(mc)^2}+\Bigl[1-\frac{(p_i\!\cdot\! p_f)}{(mc)^2}\Bigr](s_{i\alpha}-s_{f\alpha})\Bigr\}
({\color[named]{RubineRed}V^{\alpha}}{\color[named]{Blue}J^*}-{\color[named]{RubineRed}V^{\alpha*}}{\color[named]{Blue}J})+\\
+\frac{(p_f\!\cdot\! s_i)s_{f\alpha}-(p_i\!\cdot\! s_f)s_{i\alpha}+\bigl[1+(s_i\!\cdot\! s_f)\bigr](p_{i\alpha}-p_{f\alpha})}{mc}
({\color[named]{YellowOrange}W^\alpha}{\color[named]{Blue}J^*}+{\color[named]{YellowOrange}W^{\alpha*}}{\color[named]{Blue}J}
+i\frac{\varsigma_{i\alpha\beta}p_f^\beta+\varsigma_{f\alpha\beta}p_i^\beta}{mc}
({\color[named]{YellowOrange}W^\alpha}{\color[named]{Blue}J^*}-{\color[named]{YellowOrange}W^{\alpha*}}{\color[named]{Blue}J})+
\end{gather}
\begin{gather}
+\Bigl[\varsigma_{f\mu\lambda}\varsigma_{i\nu.}^{.\lambda}+s_{f\mu} s_{i\nu}-\frac{p_{f\mu}p_{i\nu}}{(mc)^2}\Bigr]
\frac{\varepsilon^{\mu\nu\alpha\beta}}{2}({\color[named]{ForestGreen}F_{\alpha\beta}}{\color[named]{Blue}J^*}
+{\color[named]{ForestGreen}F^*_{\alpha\beta}}{\color[named]{Blue}J})+
i(s_f^\alpha-s_i^\alpha)\frac{p_f^\beta-p_i^\beta}{mc}
({\color[named]{ForestGreen}F_{\alpha\beta}}{\color[named]{Blue}J^*}-{\color[named]{ForestGreen}F^*_{\alpha\beta}}{\color[named]{Blue}J})+\\
+\Bigl[(\varsigma_{i\lambda\alpha}\tilde{\varsigma}^{\lambda\,.}_{f.\beta}-\varsigma_{f\lambda\beta}\tilde{\varsigma}^{\lambda\,.}_{i.\alpha})-
\varepsilon_{\alpha\beta\mu\nu}\bigl(\frac{p^\mu_{f}}{mc}\frac{p^\nu_{i}}{mc}+s^\mu_{f}s^\nu_{i}\bigr)\Bigr]
({\color[named]{RubineRed}V^\alpha}{\color[named]{Orange}W^{\beta*}}+{\color[named]{RubineRed}V^{\alpha*}}{\color[named]{Orange}W^\beta})+\nonumber\\
+i\frac{(p_{i\alpha}+p_{f\alpha})(s_{i\beta}+s_{f\beta})-(p_{f\beta}-p_{i\beta})(s_{f\alpha}-s_{i\alpha})
-[(p_f\!\cdot\!s_i)+(p_i\!\cdot\! s_f)]\eta_{\alpha\beta}}{mc}({\color[named]{RubineRed}V^\alpha}{\color[named]{YellowOrange}W^{\beta*}}-
{\color[named]{RubineRed}V^{\alpha*}}{\color[named]{YellowOrange}W^{\beta}})+\\
+\Bigl\{\frac{(p_f^\alpha-p_i^\alpha)\Bigl[\bigl[1-(s_i\!\cdot\!s_f)\bigr]\eta^{\beta\mu}+(s_i^\beta s_f^\mu+s_f^\beta s_i^\mu)\Bigr]}{mc}+
\frac{\bigl[s_f^\alpha(s_i\!\cdot\!p_f)-s_i^\alpha(s_f\!\cdot\!p_i)\bigr]\eta^{\beta\mu}+s_i^\alpha s_f^\beta(p_i^\mu+p_f^\mu)}{mc}\Bigr\}
({\color[named]{ForestGreen}{F}_{\alpha\beta}}{\color[named]{RubineRed}V_\mu^*}+
{\color[named]{ForestGreen}{F}^*_{\alpha\beta}}{\color[named]{RubineRed}V_\mu})+\nonumber\\
+i\Bigl[\frac{(p_{i\lambda}\varsigma_f^{\alpha\mu}+p_{f\lambda}\varsigma_i^{\alpha\mu})\eta^{\beta\lambda}-
(p_{i\lambda}\varsigma_f^{\alpha\lambda}+p_{f\lambda}\varsigma_i^{\alpha\lambda})\eta^{\beta\mu}}{mc}+
\frac{\varsigma_i^{\alpha\beta}p_f^\mu+\varsigma_f^{\alpha\beta}p_i^\mu}{2mc}+
\frac{\varepsilon^{\alpha\beta\mu\lambda}}{2}(s_{i\lambda}+s_{f\lambda})\Bigr]
({\color[named]{ForestGreen}F_{\alpha\beta}}{\color[named]{RubineRed}V^*_\mu}
-{\color[named]{ForestGreen}F^*_{\alpha\beta}}{\color[named]{RubineRed}V_\mu})+\\
+\Bigl[(\varsigma_i^{\alpha\lambda}s_{f\lambda}+\varsigma_f^{\alpha\lambda}s_{i\lambda})\eta^{\beta\mu}-
(\varsigma_i^{\alpha\mu}s_{f\lambda}+\varsigma_f^{\alpha\mu}s_{i\lambda})\eta^{\beta\lambda}-
\frac{\varsigma_i^{\alpha\beta}s_f^\mu+\varsigma_f^{\alpha\beta}s_i^\mu}{2}-
\frac{\varepsilon^{\alpha\beta\mu\lambda}}{2}\frac{p_{i\lambda}+p_{f\lambda}}{mc}\Bigr]
({\color[named]{ForestGreen}F_{\alpha\beta}}{\color[named]{YellowOrange}W^*_\mu}
+{\color[named]{ForestGreen}F^*_{\alpha\beta}}{\color[named]{YellowOrange}W_\mu})+\nonumber\\
+i\Bigl\{(s_f^\alpha-s_i^\alpha)\Bigl[\bigl[1+\frac{(p_i\!\cdot\!p_f)}{(mc)^2}\bigr]\eta^{\beta\mu}-
\frac{p_i^\beta p_f^\mu+p_f^\beta p_i^\mu}{(mc)^2}\Bigr]
+\frac{[p_i^\alpha(s_i\!\cdot\!p_f)-p_f^\alpha(s_f\!\cdot\!p_i)]\eta^{\beta\mu}+p_f^\alpha p_i^\beta(s_i^\mu+s_f^\mu)}{(mc)^2}\Bigr\}
({\color[named]{ForestGreen}F_{\alpha\beta}}{\color[named]{YellowOrange}W^*_\mu}
-{\color[named]{ForestGreen}F^*_{\alpha\beta}}{\color[named]{YellowOrange}W_\mu}).
\end{gather}
\end{widetext}
Here $\tilde{\varsigma}_{\alpha\beta}$ --- dual to the $\varsigma_{\alpha\beta}$ tensor
\begin{eqnarray}
\tilde{\varsigma}_{\alpha\beta}=\frac{1}{2}\varepsilon_{\alpha\beta\mu\nu}\varsigma^{\mu\nu}=
\frac{1}{mc}(s_\alpha p_\beta-s_\beta p_\alpha).
\end{eqnarray}
Also usual designation for the dot product is used $(a\!\cdot\!b)=a^\alpha b_\alpha$.
Expression (13)-(27) determines the amplitude square $|\bar{u}_fMu_i|^2$ for any quantum electrodynamics process with polarized particles.
It has obviously Lorentz's covariant form.  This expression helps to get rid of the time-consuming necessity of trace matrices products calculations
for different processes.
Results of such calculations are already included into (13)-(27). The only thing we need to do is to substitute  specific coefficients
 $I$, $V_\alpha$, $W_\alpha$, $F_{\alpha\beta}$, $J$ for the interaction matrix $M$ into (13)-(27).
It is essentially reducing and simplifing calculations especially for the polarized particles.
Expression (13)-(27) is very cumbersome.
This is our price for it's universality.
Note that for the specific processes many of the quantities $I$, $V_\alpha$, $W_\alpha$, $F_{\alpha\beta}$, $J$ are zero so that
only some fragments of the (13)-(27) are used.
 These fragments are marked by different numbers in (13)-(27).  In each particular case expression (13)-(27) becomes
 much simpler.  As an example of such simplification let us use (13)-(27) for calculation of $|\bar{u}_fMu_i|^2$ for an electron-muon
 collision.

\section{Electron-muon collision}
The electron-muon system transaction probability per unit time from the initial state to the final state can be calculated in the usual way:
\begin{gather}
w_{fi}=\frac{|S_{fi}|^2}{T}=\nonumber\\
=(2\alpha\hbar)^2\frac{|\bar{u}_f\gamma^\alpha u_i\bar{U}_f\gamma_\alpha U_i|^2}{|{p}_f-{p}_i|^4}
\frac{c^2}{V^2}\frac{(2\pi\hbar)^3}{p_{i0}p_{f0}P_{i0}P_{f0}}\rho(E).
\end{gather}
Here $\alpha$ --- fine structure constant, $V$ --- normalization volume, which contains one electron and one muon,
$\rho(E)$ --- final states density of the system with total energy $E=c\mathcal{P}_0=c(p_{i0}+P_{i0})=c(p_{f0}+P_{f0})$
and 3D impulse $\bm{\mathcal{P}}=\bm{p}_f+\bm{P}_f$:
\begin{equation}
\rho(E)=\frac{V}{c(2\pi\hbar)^3}\frac{p_{f0}P_{f0}p_{f}^3}{\mathcal{P}_0p_f^2-p_{f0}(\bm{\mathcal{P}}\cdot\bm{p}_f)}d\Omega,
\end{equation}
$d\Omega$ --- solid angle, through the which electron is scattered. In (29)-(30) for electron (muon) quantities lower-case (upper-case) letters are used.
Expression $|\bar{u}_f\gamma^\alpha u_i\bar{U}_f\gamma_\alpha U_i|^2$ can be written as $|\bar{u}_f\gamma^\alpha u_iV_\alpha|^2$ or $|v^\alpha\bar{U}_f\gamma_\alpha U_i|^2$, where $V_\alpha=\bar{U}_f\gamma_\alpha U_i$  or $v^\alpha=\bar{u}_f\gamma^\alpha u_i$.  Amplitude square $|\bar{u}_f\gamma^\alpha u_i\bar{U}_f\gamma_\alpha U_i|^2$ can be obtained using only fragment (15) from (13)-(27).  The following quantities are zeroes  $I=0$, $W_\alpha=0$, $F_{\alpha\beta}=0$, $J=0$. Then we need to contract tensor coefficient in front of  the $V^\alpha V^\beta$, calculated for the electron, with the similar tensor coefficient calculated for the muon. Note that the real parts of these coefficients are symmetrical tensors and the imaginary parts are anti-symmetrical  tensors.  That is why we must contract them separately and add the contraction results:
\begin{widetext}
\begin{gather*}
|\bar{u}_f\gamma^\alpha u_i\bar{U}_f\gamma_\alpha U_i|^2=
\end{gather*}
\begin{gather}
=\frac{(mc)^2(Mc)^2}{16}\Bigl\{\Bigl[\bigl(s_i^\mu s_f^\nu+\frac{p_i^\mu p_f^\nu}{(mc)^2}-\varsigma_{i\lambda\,.}^{.\mu}\varsigma_f^{\lambda\nu}\bigr)
(\eta_{\mu\alpha}\eta_{\nu\beta}+\eta_{\mu\beta}\eta_{\nu\alpha})+\bigl[1-(s_i\!\cdot\!s_f)-\frac{(p_i\!\cdot\!p_f)}{(mc)^2}+
\frac{1}{2}\varsigma_{i\mu\nu}\varsigma_f^{\mu\nu}\bigr]\eta_{\alpha\beta}\Bigr]\times\nonumber\\
\Bigl[\bigl(S_{i\rho} S_{f\tau}+\frac{P_{i\rho} P_{f\tau}}{(mc)^2}-\varSigma_{i\lambda\rho}\varSigma_{f\,.\tau}^{\lambda\,.}\bigr)
(\eta^{\rho\alpha}\eta^{\tau\beta}+\eta^{\rho\beta}\eta^{\tau\alpha})+\bigl[1-(S_i\!\cdot\!S_f)-\frac{(P_i\!\cdot\!P_f)}{(mc)^2}+
\frac{1}{2}\varSigma_{i\mu\nu}\varSigma_f^{\mu\nu}\bigr]\eta^{\alpha\beta}\Bigr]+\nonumber\\
+2\Bigl[\frac{\bigl[(p_f-p_i)\!\cdot\!(P_f-P_i)\bigr]}{mcMc}\bigl[(s_f+s_i)\!\cdot\!(S_f+S_i)\bigr]
-\frac{\bigl[(p_f-p_i)\!\cdot\!(S_f+S_i)\bigr]}{mc}\frac{\bigl[(P_f-P_i)\!\cdot\!(s_f+s_i)\bigr]}{Mc}\Bigr]\Bigr\}.
\end{gather}
\end{widetext}
Expression (29)-(31) determines the transaction probability per unit time for the scattering of polarized electrons and muons.
For the unpolarized particles one must average (31) by the initial polarizations of the particles and summing by the final polarizations of  electrons and muons. It can be easily done in expression (31):  all terms with $s_{i,f}^{\alpha}$, $S_{i,f}^{\alpha}$, $\varsigma_{i,f}^{\alpha\beta}$, $\varSigma_{i,f}^{\alpha\beta}$ must be omitted and the result must be multiplied by 4 (2 for the electrons and 2 for the muons).
\begin{gather}
|\bar{u}_f\gamma^\alpha u_i\bar{U}_f\gamma_\alpha U_i|^2=\nonumber\\
=\frac{(mc)^2(Mc)^2}{4}\Bigl[\frac{p_{f\alpha}p_{i\beta}+p_{i\alpha}p_{f\beta}-(p_f p_i)\eta_{\alpha\beta}}{(mc)^2}
+\eta_{\alpha\beta}\Bigr]\times\nonumber\\
\Bigl[\frac{P_f^\alpha P_i^\beta+P_i^\alpha P_f^\beta-(P_f P_i)\eta^{\alpha\beta}}{(Mc)^2}
+\eta^{\alpha\beta}\Bigr]=\nonumber\\
=\frac{1}{2} \bigl[(p_iP_i)(p_fP_f)+(p_iP_f)(p_fP_i)-(Mc)^2(p_ip_f)- \nonumber\\
 -(mc)^2(P_iP_f)+2(mc)^2(Mc)^2 \bigr].
\end{gather}
If moreover the muon is at rest
\begin{gather*}
\bm{P}_i=\bm{P}_f=0,\hspace{1mm}|\bm{p}_i|=|\bm{p}_f|=p,\hspace{1mm} P_{i0}=P_{f0}=Mc,\\
p_{i0}=p_{f0}=p_0,\hspace{1mm}\mathcal{P}_0=p_{i0}+Mc,
\end{gather*}
for this case we get
\begin{gather*}
|\bar{u}_f\gamma^\alpha u_i\bar{U}_f\gamma_\alpha U_i|^2=
(Mc)^2\frac{p_{0}^2+(\bm{p}_i\cdot\bm{p}_f)+(mc)^2}{2}=\\
=(Mc)^2 p_{0}^2\Bigl(1-\frac{v^2}{c^2}\sin^2\frac{\theta}{2}\Bigr),
\end{gather*}
\begin{gather*}
\rho(E)=\frac{V}{c}\frac{ p_{0} p}{(2\pi\hbar)^3}d\Omega,\hspace{7mm}
|{p}_f-{p}_i|^4=\bigl(2p\sin\frac{\theta}{2}\bigr)^4,
\end{gather*}
\begin{gather}
w_{fi}=\frac{(\alpha\hbar)^2}{4}\frac{c}{V}
\frac{ p_{0} p}{\bigl(p\sin\frac{\theta}{2}\bigr)^4}\Bigl(1-\frac{v^2}{c^2}\sin^2\frac{\theta}{2}\Bigr)d\Omega.
\end{gather}
Here $\theta$ --- angle between $\bm{p}_i$ and $\bm{p}_f$. Divide (33) by the electron beam density $\frac{v}{V}=\frac{p}{p_0}\frac{c}{V}$ and
obtain the well-known result  --- Mott cross-section \cite{m}:
\begin{gather}
\frac{d\sigma}{d\Omega}=\frac{(\alpha\hbar)^2}{4}p_{0}^2\frac{\bigl(1-\frac{v^2}{c^2}\sin^2\frac{\theta}{2}\bigr)}{\bigl(p\sin\frac{\theta}{2}\bigr)^4}=\nonumber\\
=\frac{r_0^2}{4}\frac{\bigl(1-\frac{v^2}{c^2}\sin^2\frac{\theta}{2}\bigr)}{(\frac{v}{c}\sin\frac{\theta}{2})^4}\bigl(1-\frac{v^2}{c^2}\bigr).
\end{gather}
Here $r_0={\alpha\hbar}/{mc}$ --- classical electron radius. As we can see using expression (13)-(27) allows us to easily obtain process
cross-section without a calculation of the trace of the matrix products.

\end{document}